\documentclass[conference,10pt,a4paper,twocolumn,twoside]{IEEEtran}
\IEEEoverridecommandlockouts
\usepackage{caption}
\usepackage{subcaption}
\usepackage{cite}
\usepackage{amsmath,amssymb,amsfonts}
\usepackage{algorithmic}
\usepackage{graphicx}
\usepackage{textcomp}
\usepackage{xcolor}
\usepackage{braket}
\usepackage{comment}
\usepackage{algorithm}
\usepackage{algorithmic}

\usepackage{hyperref}

\def\BibTeX{{\rm B\kern-.05em{\sc i\kern-.025em b}\kern-.08em
    T\kern-.1667em\lower.7ex\hbox{E}\kern-.125emX}}

\begin{document}

% ----------------------------------------------------
% Title, authors and addresses
% ----------------------------------------------------
\title{The Quantum Internet: Enhancing Classical Internet Services one Qubit at a Time
\thanks{A.~S.~Cacciapuoti, J.~Illiano, S.~Koudia, and M.~Caleffi are with the \href{www.quantuminternet.it}{www.QuantumInternet.it} research group at the Dept. of Elect. Engineering and Inform. Technologies, University of Naples \textit{Federico II}, Italy. K.~Simonov is with s7 rail technology GmbH, Linz, Austria. A.~S.~Cacciapuoti and M.~Caleffi are also with the National Laboratory of Multimedia Communications, National Inter-University Consortium for Telecommunications, Italy.}
}

\author{
    \IEEEauthorblockN{Angela Sara Cacciapuoti, Jessica Illiano, Seid Koudia, Kyrylo Simonov, Marcello Caleffi} 
}

\maketitle

% ----------------------------------------------------
% Abstract and keywords
% ----------------------------------------------------
\begin{abstract}
Nowadays, the classical Internet has mainly envisioned as the underlying communication infrastructure of the Quantum Internet, aimed at providing services such as signaling and coordination messages. However, the interplay between classical and Quantum Internet is complex and its understanding is pivotal for an effective design of the Quantum Internet protocol stack. The aim of the paper is to shed the light on this interplay, by highlighting that such an interplay is indeed bidirectional rather than unidirectional. And the Quantum Internet exhibits the potential of supporting and even enhancing classical Internet functionalities.
\end{abstract}


% ----------------------------------------------------
% Sec. I
% ----------------------------------------------------
\section{Introduction}
\label{sec:1}
Quantum Internet is attracting worldwide research interest, given its potential of enabling a set of applications with no counterpart in the classical Internet~\cite{WanRahLi-22}.

Yet a misconception that may arise with the research interest is the idea of the Quantum Internet eventually replacing classical Internet. As a matter of fact, the very opposite holds. The Quantum Internet can not operate unassisted nor independently from the classical Internet~\cite{KozWehVan-22, CacCalTaf-20, PirBra-16}, but it rather -- widely and extensively -- depends on classical network functionalities and services for providing quantum network functionalities.

A pivotal example for this dependence is provided by the \textit{quantum teleportation} process, which represents one of the key communication protocols enabled by the Quantum Internet infrastructure~\cite{CacCalVan-20}. Specifically, quantum teleportation constitutes an astonishing strategy for \textit{transmitting} a qubit without the physical transfer of the particle storing the qubit. But it requires two different communication resources. One is quantum, i.e., a pair of (maximally) entangled qubits shared between source and destination. And the other is classical, i.e., a pair of bits transmitted from the source to the destination. Indeed, classical signaling is not limited to teleportation, but it rather constitutes a requirement widespread within the different quantum network tasks and functionalities~\cite{IllCalMan-22}, ranging from entanglement generation through distillation to swapping ad discussed in Section~\ref{sec:2}.

From the above, it becomes evident that the successful design of the Quantum Internet must carefully assess and account for the interdependence between classical Internet and Quantum Internet. Yet despite its fundamental role, such an interdependence is still underanalyzed, and its deep understanding represents a crucial open problem~\cite{IllCalMan-22}. 

The aim of this article is precisely to shed light on the interdependence between classical Internet and Quantum Internet, with the objective of allowing the reader:
\begin{itemize}
    \item to acknowledge the deep interplay between classical Internet and Quantum Internet;
    \item to understand that this interplay is bidirectional rather than unidirectional, with the Quantum Internet exhibiting the potential of supporting and even enhancing classical Internet functionalities;
    \item to appreciate the profound impact of this classical-quantum interplay on the design of the Quantum Internet protocol stack.
\end{itemize}

To this aim, we first substantiate the complex, bidirectional nature of the interplay between classical Internet and Quantum Internet in Section~\ref{sec:2}. Then, we analyze the potentialities of the Quantum Internet to enhance the classical Internet services in Section~\ref{sec:3}. This is corroborated through different use cases involving communication functionalities, ranging from physical through data link to network layer. Finally, we conclude the paper in Section~\ref{sec:4}.

% ----------------------------------------------------
% Sec. II
% ----------------------------------------------------
\section{Classical vs. Quantum Internet Interplay}
\label{sec:2}

\begin{figure*}
    \centering
    \begin{subfigure}[c]{0.3\textwidth}
        \centering
        \includegraphics[width=\textwidth]{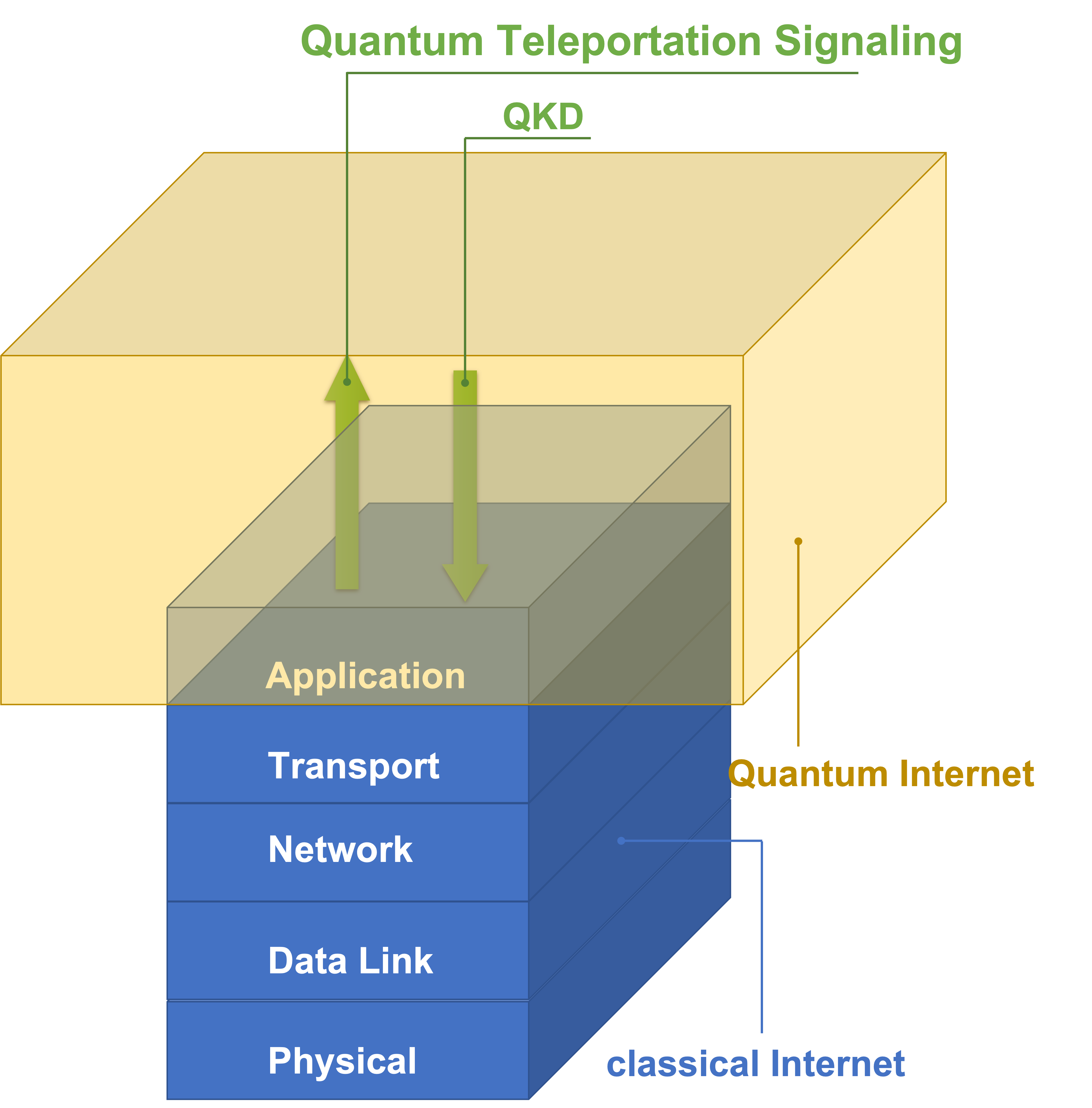}
        \caption{From the classical Internet perspective, quantum teleportation can be regarded as some sort of complex application laying on top of the classical Internet protocol stack and down-calling classical end-to-end communication services.}
		\label{fig:01.a}
    \end{subfigure}
    \hspace{0.02\textwidth}
    \begin{subfigure}[c]{0.3\textwidth}
        \centering
        \includegraphics[width=\textwidth]{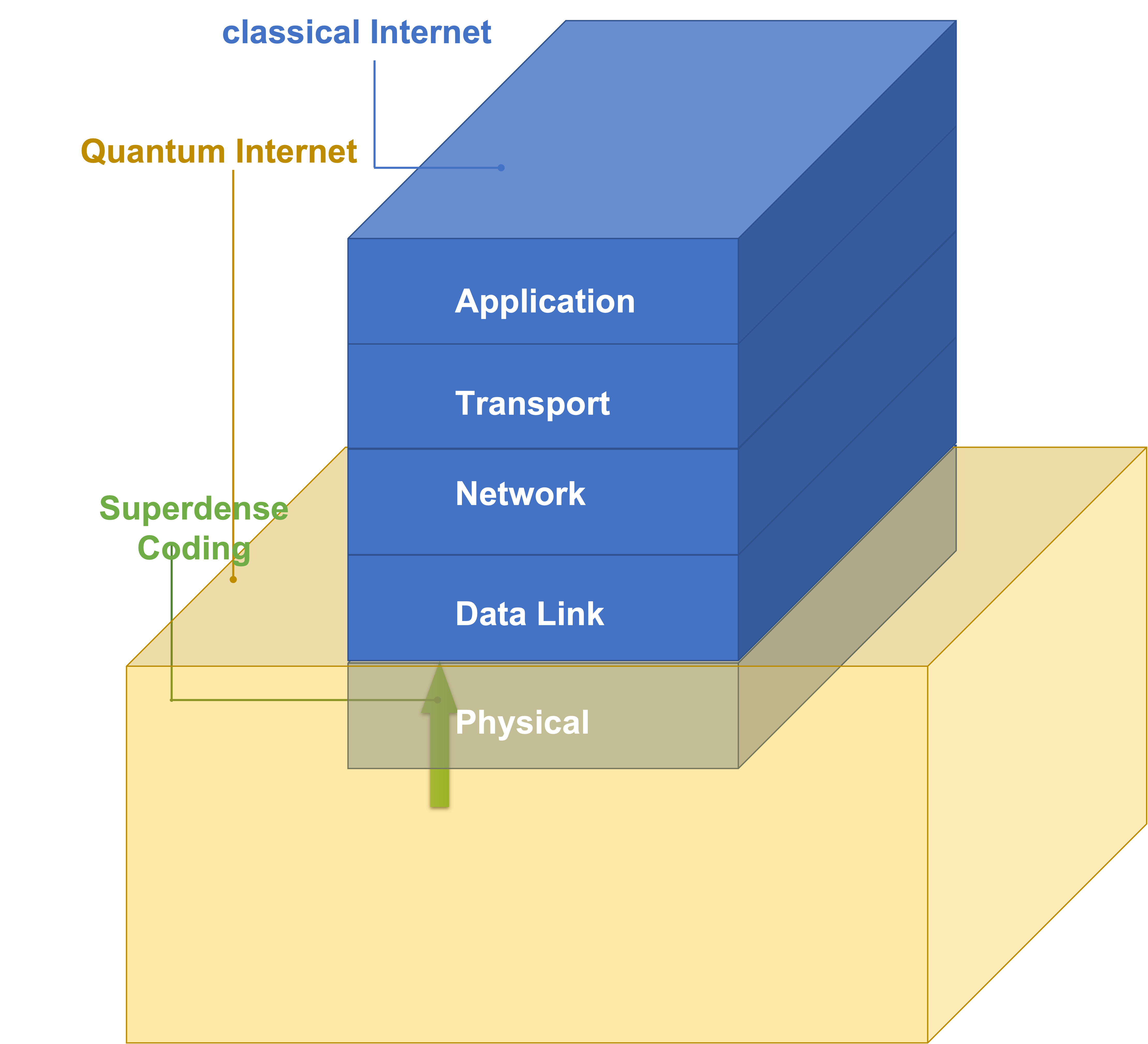}
        \caption{From the classical Internet perspective, superdense coding can be regarded as some sort of complex functionality of the classical physical layer.}
		\label{fig:01.b}
    \end{subfigure}
    \hspace{0.02\textwidth}
    \begin{subfigure}[c]{0.33\textwidth}
        \centering
        \includegraphics[width=\textwidth]{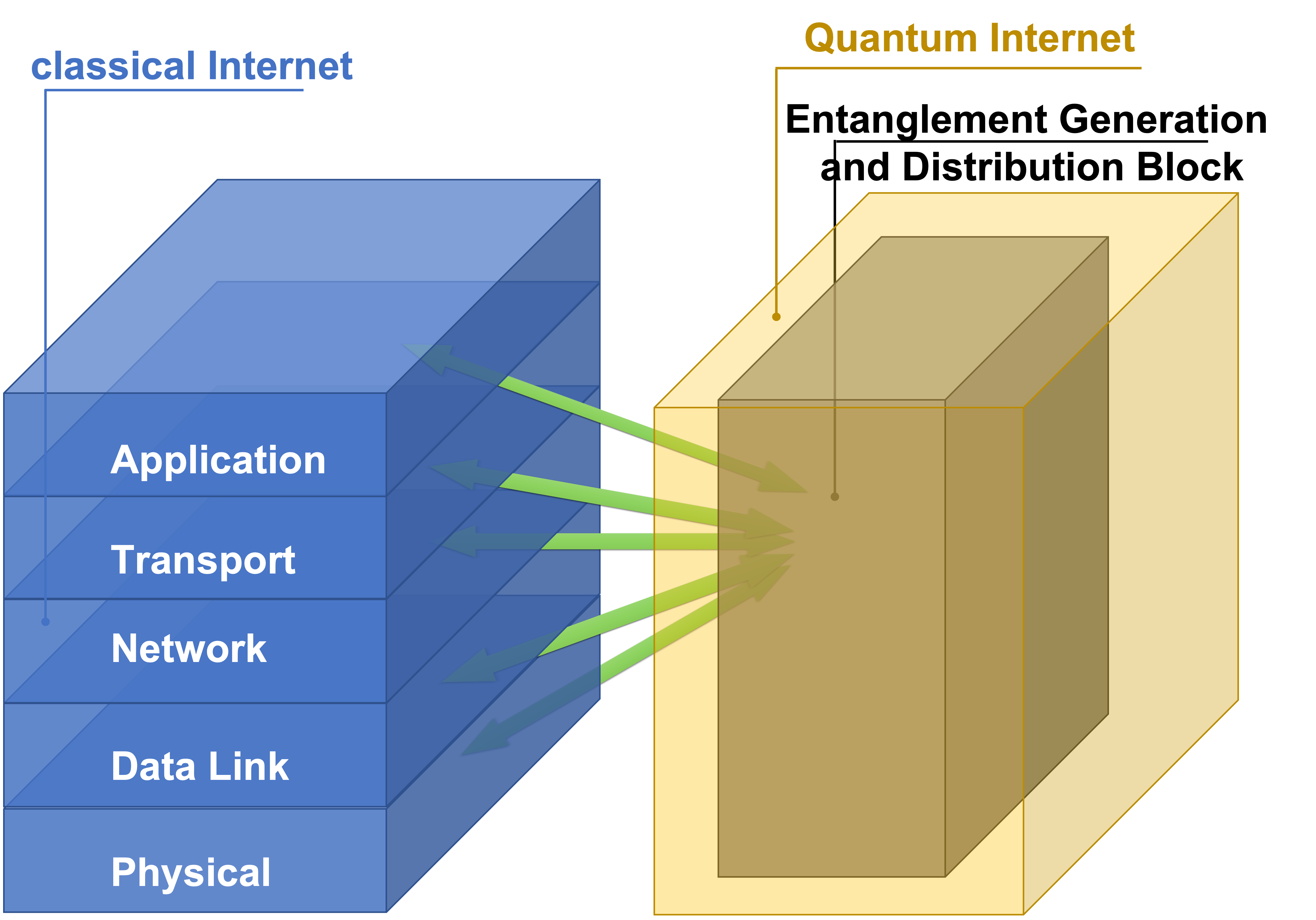}
	    \subcaption{From the classical Internet perspective, the core functionality of the Quantum Internet -- i.e., entanglement generation and distribution (grey box) as well as its manipulation -- requires the availability of multiple services from all the classical Internet protocol stack.}
		\label{fig:01.c}
    \end{subfigure}
    \caption{Interplay between classical Internet (blue boxes) and Quantum Internet (yellow box) protocol stack. The arrows represent an interaction between the two protocol stacks, with arrow direction flowing from the stack offering a service to the stack demanding such a service. The different layers constituting Quantum Internet protocol stack has been omitted given the lack of a univocal standard~\cite{IllCalMan-22}.}
    \label{Fig:01}
\end{figure*}

As highlighted, the Quantum Internet depends on the availability of classical communication and network functionalities. And, indeed, it is fairly reasonable to assume these classical services provided by the existing classical Internet. As a consequence, the interplay between classical Internet and Quantum Internet must be properly understood and modeled as a preliminary, mandatory task for the design of the protocol stack of the Quantum Internet.

In this context, classical Internet has been implicitly considered so far as an underlying communication infrastructure, aimed at providing services to the Quantum Internet. Accordingly and by oversimplifying, the Quantum Internet can been modeled as some sort of complex system laying on top of the classical Internet protocol stack -- as shown in Figure~\ref{fig:01.a} -- and interacting with the former at the application layer.

Quantum teleportation constitutes a representative case justifying such a modeling. In fact, the teleportation of a qubit is implemented with a sequence of quantum functionalities -- namely, i) entanglement generation and distribution as well as ii) local quantum operations at source and at destination -- interleaved by the transmission of a classical message. Accordingly, it seems reasonable from the classical Internet perspective -- and in agreement with the \textit{separation of concern} approach, the key principle behind OSI and TCP/IP design -- to consider quantum teleportation as some sort of complex application down-calling classical end-to-end communication services provided by the classical Internet stack.

Similar reasoning holds by considering another fundamental quantum communication protocol, namely, Quantum Key Distribution (QKD)~\cite{WanRahLi-22}. Essentially, QKD protocols fit with the former modeling where the interplay between classical and quantum occurs at the application layer of the classical Internet stack~\cite{MehMauRas-17}. However, differently from quantum teleportation, QKD provides -- rather than requests -- a service to classical Internet, by exploiting the unconventional properties of quantum systems -- such as superposition or entanglement -- to generate keys for encrypting (and decrypting) a classical message.

On the other hand, a different modeling arises from considering another popular quantum communication protocol -- namely, \textit{quantum superdense coding}~\cite{WanRahLi-22} -- that enables the transmission of two bits by ``coding'' them into a qubit, under the assumption of sender and receiver pre-sharing an entangled resource. By accounting for the specificity of superdense coding, it sounds more reasonable to envision it -- from the classical Internet perspective -- as a sort of complex functionality of the physical layer of the classical Internet protocol stack, as shown in Figure~\ref{fig:01.b}. According to this model, packets that come from the data link layer can be either encoded and then transmitted classically through the classical physical layer or be directed to a some sort of quantum super-physical layer to be encoded according to the superdense protocol. 

Indeed, while classical Internet provides quantum teleportation with classical communication functionalities, when it comes to superdense coding the opposite holds: a classical Internet functionality as bit transmission is implemented through a quantum protocol. This difference is clearly represented within Figure~\ref{Fig:01} in terms of ``relative placement'' of the Quantum Internet stack with respect to Internet stack.

As a matter of fact, the above discussion has been conducted by neglecting the fundamental communication resource of the Quantum Internet, namely, entanglement. In fact, most of the quantum protocols require, as a prerequisite, the distribution of entangled quantum states shared between source and destination~\cite{KozWehVan-22}. However, entanglement generation and distribution depend on a tight synchronization as well as on proper classical signaling exchanged between the entangled nodes~\cite{IllCalMan-22}. Indeed, classical communication services are not required only for entanglement generation: they rather constitute an essential requirement for different functionalities -- spread along the whole quantum protocol stack -- of the Quantum Internet. In this regard, it is worthwhile to point out that (together with the pivotal example of teleportation) classical signaling is mandatory for entanglement swapping~\cite{VanMet-14}. Indeed, when it comes to swapping or teleporting, signaling is not limited to neighbor nodes -- namely, host-to-host -- but they rather require end-to-end classical signaling. As a consequence, network-layer signaling among nodes belonging to different networks is required, as graphically represented in Figure~\ref{fig:01.c}.

\begin{figure}
    \centering
    \includegraphics[width=8.6cm]{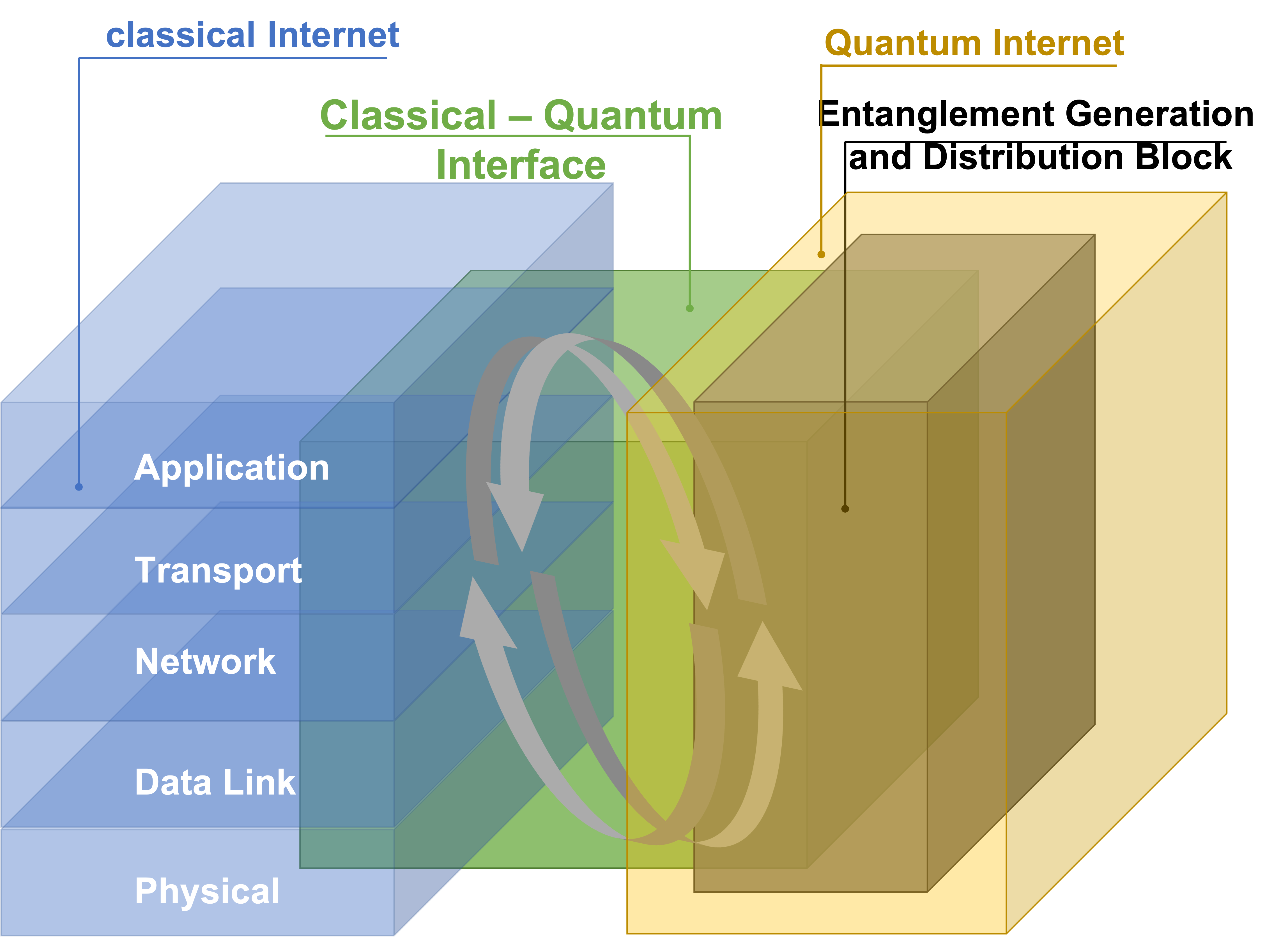}
    \caption{Classical-quantum interface (green box). The deep interplay between quantum Internet and classical Internet is not bounded to some specific layers. Conversely, it is made by a tangle of cross-layer inter-dependencies.}
    \label{fig:02}
\end{figure}

From the above, it becomes clear that the interplay between classical Internet and Quantum Internet can not be limited to a single classical-quantum interface between a classical layer offering (or requiring) some specific service to a quantum counterpart layer. But it rather requires several interactions -- likely differing in which part (quantum or classical) behaves as communication service provider -- potentially involving different layers of the classical Internet protocol stack, as graphically represented in Figure~\ref{fig:02}. In the next section, we further elaborate on this complex interactions between classical Internet and Quantum Internet, by enlarging the perspective from design challenges to opportunities.

% ----------------------------------------------------
% Sec. III
% ----------------------------------------------------
\section{Augmenting classical Internet with Quantumness}
\label{sec:3}

From the discussion above, it follows straightforward that a successful design of the Quantum Internet requires an accurate understanding of the interplay between classical and Quantum Internet. So far, we mainly highlighted the dependence of Quantum Internet on classical network functionalities spread among multiple classical layers, with two notable exceptions: QKD and superdense coding. Indeed, out of these two exceptions -- namely, quantum protocols enriching and supporting classical network functionalities rather than depending on -- only one, QKD, is attracting wide research and industrial interest due to its application potential~\cite{WanRahLi-22}. In fact, superdense coding still has not proven any practical advantage over classical bit transmission due to the challenges arising with entanglement generation and distribution.

In the following, we try to bridge this gap by discussing further examples of classical Internet functionalities that may benefit from or may be enhanced by some quantumness. To this aim, we provide the reader with three different use cases, showing how adding quantum resources to classical layers can lead to an enhancement of the classical Internet in its crucial functionalities. Starting from the lowest layer of the classical protocol stack, namely, the physical layer, in Section~\ref{sec:4.1} we discuss the birth of an exclusively bit quantum transmission service that results in a gain of the point-to-point data rate. Then, in Section~\ref{sec:4.2} we face with the core service of the data link layer, by showing that -- by enriching the classical Internet with quantumness represented by entanglement -- it is possible to solve the contention of a shared channel. Last but not least, in Section~\ref{sec:4.3}, we spread the quantum benefits to the network layer by extending the bit quantum transmission service to end-to-end communication to boost the end-to-end throughput. Furthermore, we hit the network layer with a resource with no counterpart in classical Internet, namely, the computational power of quantum distributed computing.

The aim of the following is to acknowledge that, by recalling quantum resources through quantum services, the classical Internet is enriched in its performance.

% ----------------------------------------------------
\subsection{Physical Layer}
\label{sec:4.1}

An example of classical Internet functionalities that may benefit from quantumness is bit transmission at physical layer. Specifically, in the classical Internet, the achievable data rates are upper-bounded by the physical channel capacities, with no information transmitted reliably whenever the channel exhibits zero capacity. Additionally, if the information is transmitted through a concatenation of two different channels with different capacities, the data rate is upper-bounded by the minimum of the considered capacities. Similarly, if the information is transmitted through parallel channels, the data rate is upper-bounded by the sum of the individual capacities, according to the \textit{additivity} property of the capacities.

Surprisingly, the paradigm shift from classical to quantum -- imposed by Moore's law and stimulated by Landauer: ``Information is physical'' -- comes with a whole new dazzling phenomena that overcome the aforementioned information bottlenecks. Specifically, information can be encoded in quantum carriers that propagate through their own quantum transmission links also known as \textit{quantum channels}. Interestingly, quantum channel capacities are not necessarily limited by the additivity: when quantum channels are used together for transmitting classical information, the overall capacity can be higher than the sum of the individual capacities characterizing the channels. This is known as the \textit{superadditivity phenomenon} of capacities of quantum channels~\cite{KouCacCal-21}, and it has no counterpart in the classical Internet. 

There is more to it. The shift of paradigm from classical to quantum does not mean only that the information carriers and their transmission links follow the rules of quantum mechanics. Recently, it has been shown that also the placement of quantum channels can be \textit{quantized} in order to beat some transmission limitations, which constitute major fundamental obstacles to the classical physical layer~\cite{KouCacCal-21}. Such limitations can be overcome by exploiting \textit{the quantum switch}, a device that places quantum channels in a genuinely quantum superposition of causal orders~\cite{Chi-19}. In particular, feeding the quantum switch with two channels characterized by zero-capacities activates a non-vanishing capacity, by beating the classical bottleneck inequality~\cite{KouCacCal-21}.

Stemming from the above, enriching the classical physical layer with some sort of quantumness allows the classical Internet to overcome existing data rate bounds and bottlenecks. Specifically, by exploiting the quantum switch and quantum phenomena such as superadditivity, the classical Internet can be boosted through a service provided by the Quantum Internet. This service, referred to as \textit{bit quantum transmission service} in Figure~\ref{fig:03}, offers the capability of transmitting classical bits over noisy channel in a newly quantum way. Crucially, it relies on the interplay between both Quantum and classical Internet services. In particular, the quantum switch functionality, to mitigate noise and to provide the throughput gain, requires classical communication of data. In this way, the new paradigm changes the core functionality of communication channel (namely, how the transmitted signal is affected by noise) and, correspondingly, the error correction technique at bit level. Hence, the bit quantum transmission service deeply affects the physical layer functionalities, and the ultimate result is the possibility of a remarkable enhancement of the transmission rates in the classical Internet.

% ----------------------------------------------------
\subsection{Data Link Layer}
\label{sec:4.2}

\begin{figure}
    \centering
    \includegraphics[width=8.6cm]{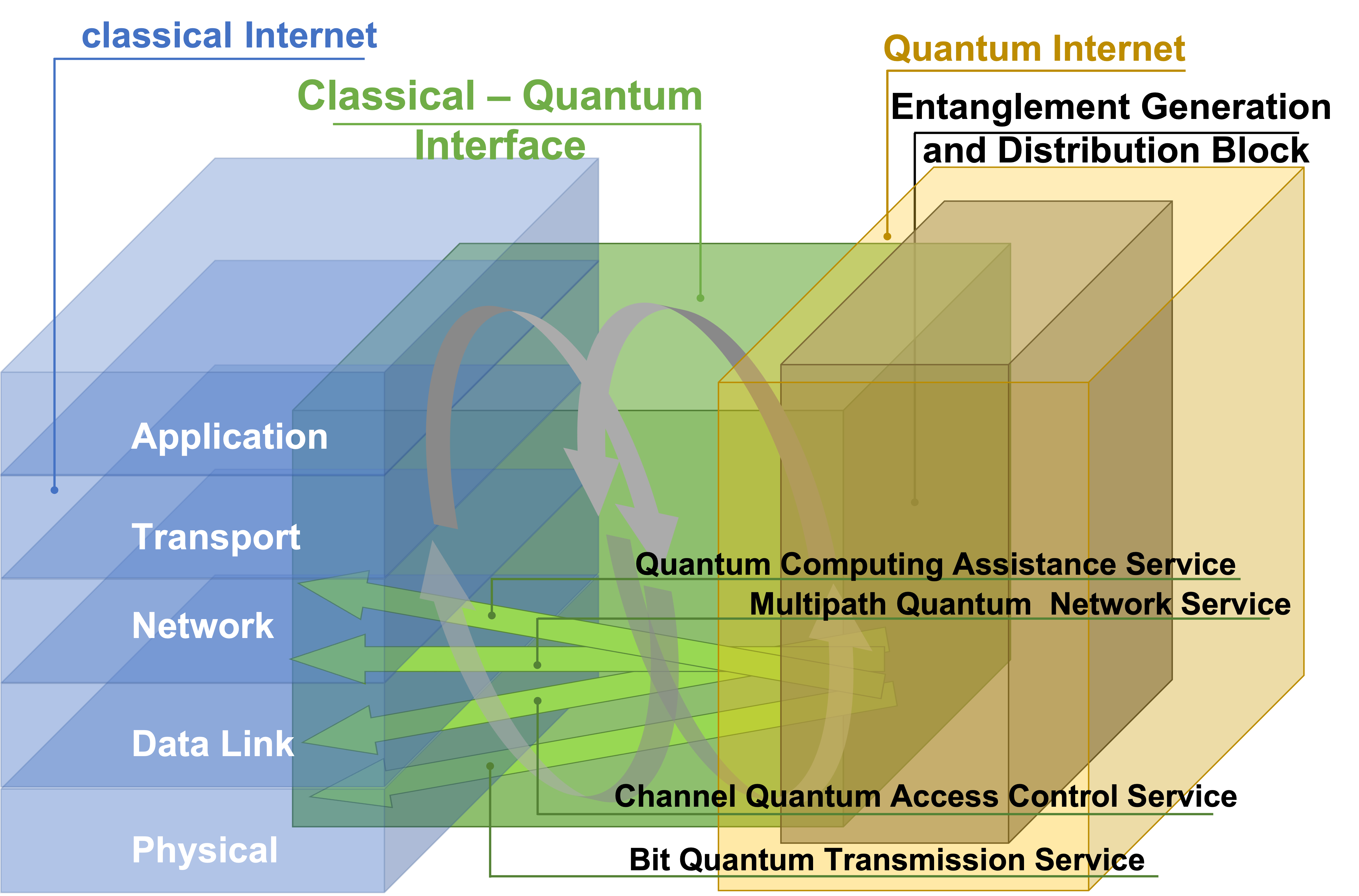}
    \caption{Examples of classical Internet functionalities that may benefit from or may be enhanced by some quantumness provided by the Quantum Internet.}
    \label{fig:03}
\end{figure}

The benefits arising from the interplay between the classical Internet and the Quantum Internet are not limited to the classical physical layer functionalities. Quantumness -- specifically, entanglement -- may offer further benefits when exploited at the data link layer of the classical Internet.

Specifically, within the classical Internet framework, communication resources such as channels are not reserved to a specific node. Conversely, they are likely shared among a set of nodes. For solving the access to such a shared resource with a distributed approach, internet devices mainly adopt carrier-sense multiple access (CSMA) protocols, and/or its variations. However, such protocols are highly affected by interference and collisions. Hence, the main data link layer functionality has never been efficiently solved.

Surprisingly, entanglement natively provides a distributed, collision-free strategy for the access to a shared classical communication resource. More into details, among the entangled states there exists a specific class of multipartite entangled states, referred to as W states, exhibiting the ability of fairly and randomly electing a leader among a set of nodes~\cite{Pan-05}. Specifically, let us consider a set of $n$ nodes, each sharing a qubit being entangled in an $n$-qubit W state, that must coordinate each other to access a classical communication channel. Each of the involved nodes simply performs a local measurement of its W-state qubit. Accordingly, only one node of the set observes the outcome $1$, whereas the remaining nodes observe the outcome $0$. Crucially, each node can observe the outcome $1$ with the same probability. 

The aforementioned procedure solves the contention problem within the classical Internet without the need of classical signaling or any form of subsequent coordination among the involved nodes. Indeed, the measurement outcome $1$ corresponds to the node allowed to use the communication resource, namely, only the elected leader can transmit on the shared channel. Differently, by observing the outcome $0$ the nodes are aware of the unavailability of the channel and are not allowed to transmit. Hence, by enriching the classical nodes with quantum resources, namely, entangled qubits, with a single operation, i.e., a local measurement, the resource contention is solved. Correspondingly, the interplay between classical Internet and the Quantum Internet enriches the former. Remarkably, this strategy is not affected by interference and it is robust with the respect to the hidden node problem. A side result worthwhile to mention is the privacy of the leader identity; indeed, among the set of nodes, only the elected leader is aware of having won the resource access.

The \textit{channel quantum access control} described above and depicted in Figure~\ref{fig:03}, entails several interaction between the classical network protocol stack and the Quantum Internet protocol stack. Clearly, it interacts with the classical data link layer. Furthermore, it relies on some sort of quantum physical connectivity for entanglement generation and distribution. Finally, it also exploits the classical physical layer, at the very least for the actual access to the channel.

% ----------------------------------------------------
\subsection{Network Layer}
\label{sec:4.3}

To continue the analysis of the possibilities arising from the interplay between the classical Internet and the Quantum Internet, we now extend the discussion about quantum benefits from point-to-point to end-to-end communications. 

Main functionalities of the classical network layer provide the necessary tools to transfer data packets from a source to destination belonging to different networks. Interestingly, a classical network architecture harnessing quantum effects allows one to route classical information encoded in quantum carriers through multiple quantum trajectories simultaneously, while preserving the quantum coherence of the quantum packet. Specifically, a point-to-point link that could not be used to transmit the information because of severe noise can turn into an effective communication channel with this genuinely quantum tool described in Sec.~\ref{sec:4.1}. This in turn affects the communications among nodes belonging to different networks, which -- by exploiting network layer functionalities (such as routing and forwarding) -- rely on point-to-point links to deliver the packet to the destination node. 

While this quantum multipath routing shares similarities with the known classical multipath routing technique, namely, fault tolerance and increased network bandwidth, it still possesses certain genuinely quantum characteristics. Indeed, \textit{the quantum path}, or \textit{quantum trajectory}, does not rely on the ability of duplicating the data packets and sending each redundant copy through different paths: this is an imposed limitation of no-cloning theorem of quantum information~\cite{CacCalVan-20}. Instead, the quantum path is manifested by a quantum superposition between multiple paths that allows transmission of a \textit{single} classical packet simultaneously through them. In other words, when quantum trajectories are allowed, the same packet is delivered via different sets of intermediate nodes and point-to-point links that exhibit different quality of service. As a result, properly assessing the quantumness with the design of a \textit{multipath quantum network service} would significantly assist the classical Internet to enhance the performance of the network layer through end-to-end paths with no classical counterpart. As a result, the interplay between the classical Internet and the Quantum Internet enriches the classical network layer.

Similarly to the \textit{channel quantum access protocol}, the \textit{multipath quantum network service} interacts with the classical network layer. From the Quantum Internet perspective, this service relies on coherent control functionalities that, in turn, depend on entanglement generation and distribution. Additionally, multipath quantum network service exploits the bit quantum transmission service in order to obtain information on quantum trajectories over point-to-point links. Accordingly and as represented in Figure~\ref{fig:03}, the interface between the classical Internet and the Quantum Internet should act as a global interface that goes beyond the separation of concerns.

Clearly, each classical Internet node only holds a partial knowledge of the network. Hence, the input data required to process the multipath quantum network service should be gathered and distributed so to converge at a given node. Then, the node must exploit computational resources for processing such a (network) topological knowledge. However, this processing can be performed by exploiting again the Quantum Internet infrastructure through the distributed quantum computing paradigm~\cite{CuoCalCac-20, CalChaCuo-20}. Indeed, the computational power of quantum computing can be used to speed-up the processing of a large amount of data~\cite{ArrDiaKer-18}, for instance routing algorithms. This framework gives birth to an additional service that the Quantum Internet can offer to the classical Internet, namely the \textit{quantum computing assistance}. As a matter of fact, it is widely known that -- among the network functionalities -- classical routing algorithms are currently difficult to solve with classical algorithms. But quantum computing can support network layer functionalities by exploiting quantum algorithms for classical routing problems~\cite{HarGamTre-21}. Indeed, as it happens for the multipath quantum network service, the information needed for processing path and routing tables could converge to the quantum computing service available at each node through the distributed quantum computing paradigm.

% ----------------------------------------------------
% Sec. IV
% ----------------------------------------------------
\section{Conclusions}
\label{sec:4}

So far, classical Internet has been mainly envisioned as an underlying communication infrastructure, aimed at providing services to the Quantum Internet. Within the manuscript, we substantiated a shift from this modeling, by discussing the bidirectional nature of the interplay between classical Internet and Quantum Internet. Indeed, the Quantum Internet is capable of boosting classical Internet functionalities laying at different layers of the classical protocol stack. With this work, we aimed at highlighting a different angle on the promises of the Quantum Internet to be. However, the interplay between classical Internet and Quantum Internet represents an undiscovered field requiring a multidisciplinary collaborative effort from communications engineering and network engineering communities.

% ----------------------------------------------------
% Sec. V
% ----------------------------------------------------
\bibliographystyle{IEEEtran}
\bibliography{biblio.bib}
\begin{IEEEbiographynophoto}
{Angela Sara Cacciapuoti} (M'10, SM'16) is a professor at the University of Naples Federico II (Italy). Since July 2018 she held the national habilitation as “Full Professor” in Telecommunications Engineering. Her work has appeared in first tier IEEE journals and she has received different awards and recognitions, including the “2021 N2Women: Stars in Networking and Communications”. For the Quantum Internet topics, she is a IEEE ComSoc Distinguished Lecturer, class of 2022-2023. Angela Sara currently serves as Area Editor for IEEE Communications Letters, and as Editor/Associate Editor for the journals: IEEE Trans. on Communications, IEEE Trans. on Wireless Communications, IEEE Trans. on Quantum Engineering, IEEE Network. She was also the recipient of the 2017 Exemplary Editor Award of the IEEE Communications Letters. From 2020 to 2021, Angela Sara was the Vice-Chair of the IEEE ComSoc Women in Communications Engineering (WICE). Previously, she has been appointed as Publicity Chair of WICE. In 2016 she has been an appointed member of the IEEE ComSoc Young Professionals Standing Committee. From 2017 to 2020, she has been the Treasurer of the IEEE Women in Engineering (WIE) Affinity Group of the IEEE Italy Section. Her current research interests are mainly in quantum communications, quantum networks and quantum information processing.
\end{IEEEbiographynophoto}

\begin{IEEEbiographynophoto}
{Jessica Illiano} received the B.Sc degree in 2018 and then the M.Sc degree in 2020 both (summa cum laude) in Telecommunications Engineering from University of Naples Federico II (Italy). In 2020 she was winner of the scholarship `` Quantum Communication Protocols for Quantum Security and Quantum Internet" fully funded by TIM S.p.A. 
Since 2018, she is a member of the Quantum Internet Research Group, FLY: Future Communications Laboratory. She is pursuing a Ph.D degree in Information Technologies and Electrical Engineering at University of Naples Federico II. Currently, she is website co-chair of N2Women. 
Her research interests include quantum communications, quantum networks and quantum information processing.
\end{IEEEbiographynophoto}
\begin{IEEEbiographynophoto}
{Seid Koudia} received the B.Sc degree in fundamental physics in 2015 and the M.Sc degree in theoretical physics with distinction in 2017 from the University of Sciences and Technology Houari Boumedien (USTHB). Currently, he is pursuing a PhD degree in Quantum technologies with the Future Communications Laboratory (FLY), Department of Electrical Engineering and Information Technology (DIETI). His research interests include quantum information theory, quantum communications, quantum networks and quantum coding theory.
\end{IEEEbiographynophoto}

\begin{IEEEbiographynophoto}
{Kyrylo Simonov} received the M.Sc. degree in physics in 2014 from the Taras Shevchenko National University of Kyiv (Ukraine) with a thesis on physics of DNA and the Ph.D. degree in physics in 2018 from the University of Vienna (Austria) with a thesis on quantum foundations. Since 2018 he worked at the Faculty of Mathematics of the University of Vienna (Austria) on mathematical foundations of quantum mechanics and applications of nonstandard analysis. His research interests include quantum information theory, quantum communications, quantum foundations, quantum thermodynamics, and mathematical foundations of quantum theory.
\end{IEEEbiographynophoto}

\begin{IEEEbiographynophoto}
{Marcello Caleffi} (M'12, SM'16) received the M.S. degree with the highest score (summa cum laude) in computer science engineering from the University of Lecce, Lecce, Italy, in 2005, and the Ph.D. degree in electronic and telecommunications engineering from the University of Naples Federico II, Naples, Italy, in 2009. Currently, he is Associate professor at the DIETI Department, University of Naples Federico II. From 2010 to 2011, he was with the Broadband Wireless Networking Laboratory at Georgia Institute of Technology, Atlanta, as visiting researcher. In 2011, he was also with the NaNoNetworking Center in Catalunya (N3Cat) at the Universitat Politecnica de Catalunya (UPC), Barcelona, as visiting researcher. Since July 2018, he held the Italian national habilitation as \textit{Full Professor} in Telecommunications Engineering. His work appeared in several premier IEEE Transactions and Journals, and he received multiple awards, including \textit{best strategy} award, \textit{most downloaded article} awards and \textit{most cited article} awards. Currently, he serves as \textit{associate technical editor} for IEEE Communications Magazine and as \textit{associate editor} for IEEE Trans. on Quantum Engineering and IEEE Communications Letters. He served as Chair, TPC Chair, Session Chair, and TPC Member for several premier IEEE conferences. In 2016, he was elevated to IEEE Senior Member and in 2017 he has been appointed as Distinguished Lecturer from the \textit{IEEE Computer Society}. In December 2017, he has been elected Treasurer of the Joint \textit{IEEE VT/ComSoc Chapter Italy Section}. In December 2018, he has been appointed member of the IEEE \textit{New Initiatives Committee}.
\end{IEEEbiographynophoto}
\end{document}